\documentclass[12pt,a4paper]{article}
\usepackage{amsthm}
\theoremstyle{plain}
\usepackage{authblk}
\usepackage[english]{babel}
\usepackage{graphicx}
\usepackage[latin1]{inputenc}
\usepackage{verbatim}
\usepackage{amsfonts}
\usepackage{amsmath}
\usepackage{txfonts}
\usepackage[T1]{fontenc}
\usepackage{color}
\usepackage{epsfig}
\usepackage{natbib}
\usepackage{tikz-cd}
\usepackage{url}
\usepackage[bookmarksnumbered=true, colorlinks=true, citecolor=blue]{hyperref}
\date{}

\title{Dependence Relations in General Relativity}
\author{Antonio Vassallo}
\affil{LOGOS-BIAP\\ University of Barcelona, Department of Philosophy\\ Carrer de Montalegre, 6\\ 08001 Barcelona\\ \url{antonio.vassallo1977@gmail.com}}

\begin{document}

\maketitle

\pdfbookmark[1]{Abstract}{abstract}
\begin{abstract}
The paper discusses from a metaphysical standpoint the nature of the dependence relation underpinning the talk of mutual action between material and spatiotemporal structures in general relativity. It is shown that the standard analyses of dependence in terms of causation or grounding are ill-suited for the general relativistic context. Instead, a non-standard analytical framework in terms of structural equation modeling is exploited, which leads to the conclusion that the kind of dependence encoded in the Einstein field equations is a novel one.\\

\end{abstract}

\vspace{6mm}

\section{Introduction}
In their comprehensive textbook on general relativity, Misner, Thorne, and Wheeler encapsulated the spirit of the theory in the famous motto ``Space acts on matter, telling it how to move. In turn, matter reacts back on space, telling it how to curve'' (\citealp{27}, p. 5). This metaphor does indeed a good job in conveying the basic idea behind the Einstein field equations, which can be schematically written as (Greek indexes run from $0$ to $3$):
\begin{equation}\label{efe}
G_{\mu\nu}[g_{\mu\nu}]=\kappa T_{\mu\nu}[\Phi, g_{\mu\nu}].
\end{equation}
\eqref{efe} basically describes the non-linear coupling (through an appropriate constant $\kappa$) of the metrical field $g_{\mu\nu}$ (on which Einstein tensor $G_{\mu\nu}$ depends) with the material field(s) $\Phi$, whose physical properties are encoded in the stress-energy tensor $T_{\mu\nu}$.\footnote{Note how the stress-energy tensor itself functionally depends on the metric tensor $g_{\mu\nu}$.}

An interesting question to be asked then is: how much should we read into Misner, Thorne, and Wheeler's description of \eqref{efe}? In particular, how seriously should we take the reference to action and reaction? In everyday thinking, we are accustomed to the idea that matter acts on other matter in a straightforward sense. For example, we have no problem to accept that, under certain conditions, we can act on a door in a way that determines its opening. Thus, whether the opening of the door obtains \emph{depends} on whether it is pushed. However, things become much less clear and intuitive when we try to apply the same picture to spatiotemporal structures determining the geodesic motions of material bodies, or to massive bodies determining the geometry of spacetime. Even adopting a skeptical attitude towards the existence of spacetime does not ease the puzzlement, because then we have to explain why Misner, Thorne, and Wheeler's metaphor is in fact so powerful.

In the following, I will analyze from a metaphysical standpoint the determination relation underlying the talk of mutual action between material and spatiotemporal structures in general relativity. I will pay particular attention to the first part of the afore-mentioned metaphor, discussing in what sense spacetime determines the geodesic motion of freely-falling bodies. In the next two sections, I will review the current philosophical debate on the subject, highlighting the controversy surrounding the notion of action. In section \ref{3} I will introduce a non-standard framework for the analysis of dependence relations, which involves structural equations modeling. Finally, in section \ref{4}, I will show how such a framework is able to clarify the source of the controversy, concluding that the relation encoded in \eqref{efe} represents a novel kind of dependence.

\section{Are spatiotemporal properties causal?}\label{2}
The most straightforward reaction one can have towards Misner, Thorne, and Wheeler's motto is to take it at face value: spacetime does not metaphorically guide matter, it literally causes bodies to freely fall. This plain causal reading of spatiotemporal properties can come in different flavors. One of the most developed proposals in this sense is due to Alexander Bird (see, e.g., \citealp{384}). In a nutshell, Bird claims that spatiotemporal properties --if fundamental-- are essentially dispositional. Such a stance relies on the later Einstein's invocation of an action-reaction principle in order to account for the reciprocity encoded in the field equations of general relativity (see \citealp{438}, for a discussion of the evolution of Einstein's views on the subject). In Bird's words:
\begin{quote}
In dispositional essentialist terms, we can see that by being potential manifestations of dispositional essences, spatial and temporal properties may also have dispositional essences themselves. [T]hat perspective is precisely that endorsed by General Relativity. Each spacetime point is characterized by its dynamical properties, i.e. its disposition to affect the kinetic properties of an object at that point, captured in the gravitational field tensor at that point. The mass of each object is its disposition to change the curvature of spacetime, that is to change the dynamical properties of each spacetime point. Hence all the relevant explanatory properties in this set-up may be characterized dispositionally.\\
(\citealp{384}, p. 240)
\end{quote}

In other words, Bird argues that, according to general relativity, spacetime is recipient of change caused by matter but, given the reciprocity encoded in the Einstein field equations, that means that spacetime is able to cause a change in matter. Under this reading, spatiotemporal properties are active in a straightforward causal sense, and the kind of dependence relation among material and spatiotemporal facts underlying the Einstein field equations looks like a close relative of everyday causation.\footnote{Here we are just taking for granted that dispositions are causally relevant, cf. \citet{505} for discussion.} Or does it?

The main conceptual problem with Bird's view is that possessing a dispositional essence necessarily entails an appropriate counterfactual.\footnote{For the time being, let us just focus on Lewis-style counterfactuals. We will later consider alternative approaches to counterfactual reasoning.} In the case at hand, such a counterfactual would be something like ``were the metrical structure to undergo a variation, the mass-energy distribution would change as well''. The way we would evaluate the truth value of this counterfactual would be: take a solution of the Einstein field equations that depicts the actual situation $\langle g_{\mu\nu},T_{\mu\nu}\rangle$, feed a small perturbation of the metric tensor in the field equations, and if they give back a model with a (slightly) different $T_{\mu\nu}$ then the counterfactual is true.\footnote{Of course, given the reciprocity encoded in \eqref{efe}, the same reasoning can be repeated mutatis mutandis for the dependence of spacetime geometry on the mass-energy distribution.}

The talk of ``small perturbations'' is the general relativistic way to refer to the possible world nearest to the actual one. The problem with this way of proceeding is that, in general, it is not possible to non-arbitrarily fix the nearest possible world. To have a rough idea,\footnote{See \citet[][section 3]{487} for a formally rigorous example.} imagine we have a world with just one massive body. Can we evaluate the counterfactual ``were the body to disappear, the geometry of the universe would be everywhere flat''? Intuitively, we would say that this counterfactual is true but, in practice, we have to compare the starting model with a solution of vacuum field equations $G_{\mu\nu}[g_{\mu\nu}]=0$. For sure, Minkowski spacetime is \emph{one} of these solutions, but not the \emph{only} one. We therefore need some additional extra-dynamical argument to fix the Minkowski solution as the nearest world to the starting one.

It does not take much to realize that the issues with evaluating counterfactual changes in general relativity are a direct consequence of the dynamical nature of the spacetime structures described by this theory. To see this, just make a comparison with Newtonian mechanics. In this latter theory, the main features of spacetime are fixed both in a physical --i.e. spacetime structures are not influenced by material ones-- and a metaphysical sense --i.e. all possible worlds in which Newtonian mechanics holds feature the same spacetime structures (see \citealp{467}, for an appraisal of background spatiotemporal structures from a metaphysical perspective). As a consequence of this metaphysical rigidity, it is always possible to use Newtonian background structures as a standard against which any counterfactual change can be evaluated. \citet{487} makes a nice case:
\begin{quote}
[O]ne may be interested in the question: what would happen to the orbits of the planets in the Solar System if the sun were to vanish? Nothing simpler. Plug in [the initial-value formulation] the new values for the sun's size and mass (\emph{viz}., zero), compute the new orbits, and compare them to the original ones by using the background simultaneity and affine structures as referential framework. [...] There is no need for the \emph{ad hoc} fixing of comparison classes of systems, or for the \emph{ad hoc} fixing of methods for identifying ``the same quantity of the same system at the same place and same time, under otherwise different conditions or in otherwise different states''. It's all fixed naturally and canonically from the start.\\
(\emph{ibid.}, p. 4)
\end{quote}

Clearly, such a referential framework is not available in general relativity, thus making the evaluation of counterfactual change a rather tricky issue. The immediate consequence of the troubles with Lewisian counterfactuals in general relativity is that the Einstein field equations by themselves are unable to provide a robust enough ``causal-like'' link between spacetime and matter.

We may ask whether there can be a different approach according to which spatiotemporal properties can be given a  straightforward causal reading without resorting to Lewis-style counterfactual analysis. In other words, can we provide a sound story according to which spacetime points (or small spacetime regions)\footnote{This way of talking should not trick the reader into believing that the present discussion requires a substantivalist attitude. In fact, the relationalist can take the reference to spacetime points or regions as a verbal shortcut to address aspects of a web of spatiotemporal relations among material bodies.} perform an action on material bodies, which ``shapes'' their trajectories? If dispositions are not enough to do the job, perhaps we can base our characterization of action on something more physically tangible, such as some sort of energy-momentum transfer. After all, this seems to be the exact idea underlying, e.g., the slingshot effect exploited in astronautics. Roughly, we speak of the slingshot effect when a spacecraft passing near, say, a massive planet, gains some energy-momentum from the planet's gravitational field and, thus, accelerates. 

This well-known physical effect might seem to perfectly fit an account of causation in terms of physical processes --such as the conserved quantity approach (see, e.g., \citealp{475}). According to this view, spacetime would ``push'' material bodies into geodesic motion by transmitting a certain amount of energy-momentum to them. This would be a picture even more physically-laden than the one given in terms of dispositions. However, there are at least two reasons to regard this position as awkward if applied to geodesic motions in general relativity. The first, and most obvious, is that geodesic motions in general relativity can be regarded as ``default'' motions in a sense similar to that of inertial motions in Newtonian mechanics, i.e. the type of motions associated to \emph{isolated} (i.e. non-interacting) bodies. Hence, it is not clear at all how the notion of energy transfer can be consistently introduced in the description of geodesic motions without running into troubles. For example, would that mean that there is a ``true'' default motion that involves no energy transfer at all, which just happens to be hidden from view because of spacetime's ``push''?

Secondly, general relativity teaches us to be very careful when considering local energy-momentum exchange processes. In a nutshell, the problem is that, in general relativity, the gravitational energy-momentum is represented by a quantity --a \emph{pseudo-tensor}-- that is not invariant under coordinate transformations. This means that gravitational energy-momentum cannot be defined locally in a unique --i.e., observer-independent-- way, which in turn disrupts the general description of energy-momentum transfer processes in a given spacetime region (for a technical justification and a philosophical discussion of this problem see \citealp{386,38}). Even if this does not automatically rule out the possibility to make substantial physical sense of gravitational energy (and transfer thereof) in general relativity (see \citealp{575}, for a recent effort in this sense), still it is apparent that defending the conserved quantity approach in this context is a tremendously hard task.

The above mentioned issues are exploited in \citet{473} to come up with a straightforward anti-causal argument. The premises of the argument are that (P1) spacetime does not \emph{force} bodies to move geodesically (it just determines what the available geodesic paths are), and (P2) causal processes always involve the action of forces (at least, in fundamental physics). Then the obvious conclusion is that spacetime does not causally act on matter.

How can Livanios' argument be contrasted? We have already seen why the denial of (P1) is rather problematic, but, even taking (P1) for granted, still Livanios' argument may be attacked by denying (P2) and, more precisely, by claiming that general relativity is in fact a fundamental physical theory where the notion of cause can be divorced from that of force. To this line of reasoning, Livanios replies as follows:

\begin{quote}
Currently, there are three interpretations of what, according to GR, the influence of spacetime on matter amounts to: Some physicists and philosophers think of GR as ``geometrising'' away the gravitational force, while others think of it as showing that all spatiotemporal phenomena are expressions of the gravitational field. Finally, there is a third view according to which, in the general relativistic context, there is a conceptual identification of the gravitational field with the proper spatiotemporal geometrical structure [cf. \citet{50}]. In neither of these interpretations we have an indication that GR \emph{dissociates} the concept of the cause from forces.\\
(\citealp{512}, pp. 24-25)
\end{quote}

The last sentence above is rather questionable. While it is true that neither of these interpretations of GR \emph{dissociate} the concept of cause from forces, it is equally true that they do not \emph{associate} the concepts either. All these interpretations address is the issue of the ``arrow of reduction'' between spacetime and the gravitational field. This has nothing to do with the issue of causation per se. 

Hence, if it is true that the problem with ascribing causal efficacy to spatiotemporal properties in general relativity just stems from tacitly requiring that such properties perform the causal job by exerting a force, then it seems that here friends of causation have a chance to redeem themselves. Indeed, they can just claim that anchoring the causal talk to the presence of forces is too strong a requirement for causal dependence. This is exactly what Andreas Bartels does:

\begin{quote}
The answer to [Livanios'] objection is that, in General Relativity, metrical (affine) structure is the means by which the gravitational field couples to matter. Spacetime affects matter not by means of classical forces, but by means of its metrical (and affine) structure which brings about tidal forces producing paradigmatic causal effects like spatial deformations of material bodies.\\
(\citealp{127}, p. 2005)
\end{quote}

Bartels bases his causal thesis on a counterfactual analysis of causation alternative to the standard Lewisian analysis considered so far, namely, the \emph{interventionist} (or, \emph{manipulationist}) framework \citep[see][for an introduction to the subject]{502}.

The core idea behind the interventionist approach is that a condition $C$ causes a condition $E$ if and only if were $C$ to be subjected to an exogenous process that changes it, $E$ would change as well. More prosaically: $C$ causes $E$ if and only if by wiggling $C$, $E$ wiggles as a result. Of course, the intervention behind the wiggling of $C$ does not have to be agential: any physical process external to the system under scrutiny would do (hence the word \emph{exogenous} used above). Insisting that an intervention has to be external to the system is very important, because we want to exclude that the wiggling of $E$ is a consequence of the change in another causal link that bypasses $C$. That being said, it is not necessary to claim that an intervention has to be a possible physical process in a strict sense.\footnote{\label{foot}To be fair, this is a contentious point. Some may object that, given that a cause is basically a ``handle'' to manipulate effects, what it is to be a cause is \emph{defined} by the very nature of the associated intervention. Hence, it is quite suspicious to allow for interventions that are not realizable --at least in principle-- through a physically well-defined process. We will return to this point in section \ref{3}. For the time being, we just stress the fact that there can be manipulationist accounts stronger than the one discussed here, in the sense that they place more strict constraints on allowed interventions.} All we require is (i) to have a ``coherent conception'' of what it is to wiggle $C$, and (ii) to be able to exclude ``confounding'' cases in which the change of $E$ is not (just) due to the wiggling of $C$.\footnote{This further clarifies the sense in which an intervention on $C$ has to be \emph{external} to the system: it has to sever any confounding causal connection ``upstream'' of $C$ (in particular it must be the case that $C$ cannot affect $E$ via any other causal route than the one probed by the intervention).} Hence, according to Woodward:
\newcounter{dummy1}
\refstepcounter{dummy1}
\begin{quote}\label{woo}
This suggests that there will be a basis for claims about what will happen to $E$ under an intervention on $C$ as long as we can associate some well-defined notion of change with $C$ and as long as we have some grounds for saying what the effect, if any, on $E$ would be of changing just $C$ and nothing else.\\ 
(\citealp{561}, p. 131)
\end{quote}

In the case of general relativity, according to Bartels, this basis clearly exists, and is represented by the formal machinery of the theory. Indeed, if we model both $C$ and $E$ via some variables related by the laws of the theory, we are able to evaluate any kind of counterfactual claim whose antecedent involves a change of value in the variable modeling the condition $C$, because it is the laws of the theory themselves that provide a coherent conception of change and a non-confounding link between antecedent and consequent. Let's see how this works in the concrete case of geodesic motions.

First of all, we model material motions by timelike curves $x^\mu(\lambda)$ ($\lambda$ representing an appropriate affine parameter): these are our dependent variables. The dynamically allowed curves --more precisely, the curves representing freely-falling bodies-- are the solutions to the equations of motion of general relativity\footnote{See \citet{576} for a voice against this ``canonical'' view of geodesic motion.} (a dot indicates standard differentiation with respect to $\lambda$):
\begin{equation}\label{geo}
\ddot{x}^\mu+\Gamma^{\mu}_{\phantom{\mu}\nu\sigma}\dot{x}^\nu\dot{x}^\sigma=0.
\end{equation}
Equations \eqref{geo} follow directly from \eqref{efe} (see, e.g., \citealp{27}, section 20.6, for a formal derivation) and represent the mathematical codification of the fact that freely-falling (i.e. non-interacting) bodies move along the ``straight lines'' of the geometry encoded in $g_{\mu\nu}$. The measure of the ``straightness'' of a curve is provided by a mathematical device called \emph{connection} (usually symbolized by the covariant derivative operator $\boldsymbol{\nabla}$), whose components are the so-called \emph{Christoffel symbols} $\Gamma^{\mu}_{\phantom{\mu}\nu\sigma}$ appearing in \eqref{geo}. These Christoffel symbols are derived from the metric tensor through the relation
\begin{equation}\label{cris}
\Gamma^{\mu}_{\phantom{\mu}\nu\sigma}=\frac{1}{2}g^{\rho\mu} \big(g_{\sigma\rho,\nu} + g_{\rho\nu,\sigma} - g_{\nu\sigma,\rho}\big),\footnote{More precisely, this relation holds for a very particular kind of connection, called \emph{Levi-Civita}. I will come back to this point in section \ref{4}.}
\end{equation}
where the comma indicates standard differentiation with respect to the subsequent index.

With this machinery in place, we proceed by indicating the exogenous variable, i.e. the one to be wiggled. In our case, this is clearly the metric tensor $g_{\mu\nu}$. Let's now consider a certain solution of the field equations $\langle g_{\mu\nu},T_{\mu\nu}\rangle$ (plus related boundary conditions, if any). Such a model will have a set of geodesic motions $\{x_{i}^\mu(\lambda)\}_{i\in\mathbb{N}}$ given by the geodesic equation. The wiggling of $g_{\mu\nu}$ here amounts to selecting a different metric $\tilde{g}_{\mu\nu}$ which is dynamically allowed, i.e., such that it is part of another model $\langle \tilde{g}_{\mu\nu},\widetilde{T}_{\mu\nu}\rangle$. We then feed $\tilde{g}_{\mu\nu}$ into the formula \eqref{cris} for the coefficients of the connection and, consequently, in the geodesic equation. Of course, the new set of solutions $\{\tilde{x}_{i}^\mu(\lambda)\}_{i\in\mathbb{N}}$ of \eqref{geo} will be different from the starting one $\{x_{i}^\mu(\lambda)\}_{i\in\mathbb{N}}$. Hence, by intervening --in the loose sense justified above-- on the geometry of spacetime, we changed the geodesic motions of material bodies. The causal verdict is thus positive: the geometrical properties of spacetime do cause bodies to move with geodesic motion. Note how it is the theory itself that provides the conceptual apparatus needed to make clear (i) what it means to change the cause $C$, and (ii) how the depending condition $E$ would change as a direct and exclusive consequence of the wiggling $C$.

By basing his causal analysis on interventionism, Bartels is able to defuse the objection that a causation-friendly treatment of spacetime in general relativity has to  imply or require that spacetime literally pushes material bodies to move geodesically. In fact, the manipulationist framework does not seek to reduce the notion of cause to any more fundamental notion (let alone that of force, or energy-momentum exchange). Of course, the fact that the interventionist account of causation is non-reductive is not a problem in the context of our discussion, since we are interested in assessing whether \emph{there is} a causal link, and not in giving a detailed conceptual account of what causation is. 

How does Bartels' proposal fare with respect to the issues with counterfactuals evaluation in general relativity? Unfortunately for friends of causation, the answer is: not that well. In fact, the manipulationist analysis sketched above does not work on ``small portions'' of a model, that is, we cannot define surgical interventions that affect only a condition local $C$ while leaving everything else untouched (which, again, is a key requirement if we want to avoid confounding cases).

In the above example, the intervention works insofar as the geometry --that is, a diffeomorphic class of metric tensors-- over the \emph{whole} model is changed: it makes no sense to ask what would have happened if we changed it just in a neighborhood of a point, leaving everything else unchanged. To see why it is so, let's assume that the model is well-behaved enough to admit a $3+1$ decomposition of the dynamics. What we would do to represent this local intervention would be to take the values of the appropriate variables on an initial Cauchy $3$-surface, make a ``small deformation'' (that is, changing the value of the metric variables) on a small neighborhood of the surface, and let the dynamics evolve the surface to subsequent ones. 

The problem is that, in this formulation, the Einstein field equations represent a geometrical constraint over the sequence of $3$-surfaces embedded in a $4$-dimensional model. Therefore, deforming --even infinitesimally!-- the geometry of the initial $3$-surface would most likely violate the embedding constraints and, hence, the field equations themselves. \citet{574} give a detailed technical justification for why it is so. Very roughly speaking, the geometrical constraints can be cast as a set of non-linear and elliptic differential equations. The non-linearity implies not only that a combination of solutions of the system of equations does not constitute a solution by itself, but also that adding a ``small deformation'' to a solution will not give in general a new solution. The ellipticity of the system, on the other hand, points at the fact that the equations do not strictly speaking describe any dynamical evolution for the initial data but, rather, they encode global constraints for such data: this implies that the value taken on by a solution at a point generally depends on all the initial data over a Cauchy surface. This is why it is extremely difficult to define a surgical intervention without ``messing up'' the whole model.

To sum up, if an intervention is a dynamically allowed change in the value of a variable, then interventions can mostly be ``model-wide'', and causation loses any interesting local connotation, i.e. setting aside few extremely favorable cases, we cannot counterfactually test any causal link that would involve a surgical intervention on a small part of the model (this problem is also discussed in the context of frame-dragging effects in \citealp[][section 3]{504}). This is awkward not only from an ontological perspective, but also from an explanatory one. For example, we would be forced to say that the precession of Mercury's perihelion is not an effect just due to the surrounding spacetime geometry, but to the whole spacetime geometry of the solar system. 

In short, buying into this type of ``causal non-locality'' would push towards a very strong type of holistic metaphysics, and would also imply that any explanation of local state of affairs that mentions only the immediate spatiotemporal surroundings of such states of affairs would be incomplete. This is quite a high price to be paid to ascribe a causal nature to spatiotemporal properties. So, all things considered, switching to an interventionist counterfactual approach is not the panacea for defenders of the causal nature of spatiotemporal and material influences in general.

To conclude our review of the literature discussing the causal nature of spatiotemporal properties in general relativity, let us consider an anti-causal argument put forward  in \citet{387}. The gist of the argument is that being committed to a causally-friendly view of spatiotemporal properties entails a commitment to physically opaque instances of causation.

Katzav starts by noticing that a cause can bring about a non-occurrence in a physically acceptable sense only as a result of the inhibition of some endeavor. For example, imagine that a skinny hiker encounters a huge boulder on his path and unsuccessfully tries to push it out of his way. In this case, we may say that the non-occurrence of any boulder's movement is caused by its huge mass. A physicist would nicely describe such a scenario in terms of Newton's second law as an interaction between the hiker and the boulder, in which the boulder pushes the ground ``harder'' than the hiker pushes the boulder. Thus, here, the mass of the rock does an active job in preventing something --i.e. its sudden movement-- from happening. However, it would be physically awkward to claim that the huge mass of the boulder causes the non-occurrence of its sudden acceleration when no external force is exerted on it. In this second case, it does not seem right to say that the rock's mass \emph{does the job} of preventing it from accelerating. In fact, a physicist would explain the absence of this kind of motion by appealing to a non-causal principle such as Newton's first law. Insisting that, also in this case, mass plays an active causal role would clearly entail a commitment to spooky instances of causation.

The next step in Katzav's argument is to translate this line of reasoning to geodesic motions in general relativity. Thus, the question is: can we say that the geometrical properties of spacetime \emph{cause} free bodies to move geodesically? Having in mind Katzav's criterion for physically acceptable instances of causation, we might be tempted to say that, yes, spacetime causes geodesic motions to happen by countering the bodies' disposition to move non-geodesically. 

However, such a picture is untenable for two reasons. First, similarly to the energy-transfer case discussed earlier, it would seem that now the ``natural'' state of motion of a body is opposed by spatiotemporal structures, and geodesic motions are the result of such an opposing influence. This is, of course, not the story told by general relativity, according to which geodesic motion is an instance of persistence as is through motion --i.e. absence of change in any relevant physical respect throughout the trajectory. Secondly, the explanation of geodesic motions in this picture would look like the enactment of a cosmic conspiracy according to which the true nature of free motions is hidden from our view. 

If, instead, we admit that geodesic motions are not an effect of countering hidden true motions, then the causal efficacy of spatiotemporal properties is called into question as unfit to back up the physical explanation of the phenomenon. Indeed, the most we can say in this case is that spacetime causes the \emph{absence} of inexplicable changes in the trajectories of freely-falling bodies. For sure, this is not a ``physics-friendly'' instance of causation.

Katzav's anti-causal argument is probably the most contentious among the ones reviewed in this section. What is most puzzling about Katzav's argument is the premise that a cause can bring about a non-occurrence in a physically
acceptable sense \emph{only} as a result of the inhibition of some endeavor. Katzav here is clearly brushing aside alternative --and equally acceptable-- instances. For example, spacetime might cause the non-occurrence of a certain trajectory $x^{\mu}(\lambda)$ by bringing about another trajectory $\tilde{x}^{\mu}(\lambda)$ that is incompatible with $x^{\mu}(\lambda)$.\footnote{I am grateful to an anonymous referee for suggesting this objection.} Given that much of  Katzav's argument depends on this premise, it looks like he is providing the least cogent anti-causal case of the batch presented in this section. However, this is just a cold comfort for the fans of causation in general relativity, given the much more troublesome issues discussed earlier --in particular, the problem of meaningfully define counterfactual change in Lewis-style analysis, and the issue with non-local interventions in the manipulationist approach.

\section{Are spatiotemporal properties non-causal?}\label{2bis}

So far, we have discussed in detail why and how general relativity does not represent a hospitable environment for fans of causation. Consequently, skeptics towards causation seem to gain the upper hand in the debate. Or do they? Otherwise said, can causal skeptics come up with a convincing metaphysical characterization of the spacetime/matter-dependence?

Let's go back to Livanios' work, and see his own proposal:
\begin{quote}
[A] more cautious analysis points out another interpretation, according to which the dynamical character of space-time allows the \emph{trans-world} variation of space-time structure in such a way that the Einstein equations hold. In such an approach, we look upon the Einstein equations as giving a law-like consistency constraint upon the joint
features (space-time structure and mass-energy distribution) of any (physically) possible world.\\
(\citealp{473}, p. 389)
\end{quote}

Judging from the above quotation, it seems that Livanios has in mind some sort of ``physical possibility'' relation, e.g., the geometry of a certain spacetime region $A$ determines (through the mediating role of \eqref{efe}) the ``physicality'' of a certain trajectory $x^{\mu}(\lambda)$ in $A$. We immediately see the problem with this proposal: it assigns an inherently modal nature to the dependence relation underpinning the Einstein field equations, and we already know that modality in general relativity is a tricky issue. Indeed, such a relation (again, through the mediating role of \eqref{efe}), should support counterfactuals like ``if the geometry of $A$ were different, it would be physically impossible for a freely falling object to move along $x^{\mu}(\lambda)$ in $A$''. In evaluating such a counterfactual, we would face the very same troubles already encountered before. As already pointed out in section \ref{2}, because of the dynamical character of spacetime, there is no objective trans-world standard against which we can assess any notion of counterfactual variation. So there is no straightforward way to individuate the ``changed version'' of $A$ in a possible world near to the starting one (to catch a glimpse of the debate on trans-world individuation of spacetime points and regions, see \citealp[][chapter 9, section 14]{118}).

The immediate reaction to my criticism is that it is unwarranted and stems from a misguided reading of Livanios' passage. In fact, Livanios' quotation should be intended as stating that this physical possibility relation is just a ``model-building'' relation that couples metric tensors with stress-energy tensors in accordance with \eqref{efe}. If that is the case, then all that we need is that only worlds in which the Einstein field equations hold are physically possible. In particular, we do not need to say anything about the evaluation of counterfactual change, possible worlds vicinity, or anything like that.

Let's accept for the sake of the argument that this is the case. Then a new substantial issue presents itself: Livanios' possibility relation becomes trivial to the point of mere analyticity, thus morphing into some sort of conceptual necessity relation. The main trouble with the relation of conceptual necessity is that it usually supports very feeble explanations. For example, once we define what a bachelor is, it becomes matter of conceptual necessity that ``$X$ is a bachelor'' obtains whenever ``$X$ is an unmarried man'' obtains. However, the analytical link between these two facts makes an explanation of the form ``$X$ is a bachelor because $X$ is an unmarried man'' rather uninformative. In a similar vein, an explanation of the form ``such and such geometry is coupled to such and such mass-energy distribution because the Einstein field equations hold'' is not very informative. The most that we can learn from such an explanation is that the explanandum is not merely an accident, in the sense that it is a particular instance of the law-like generalization constituting the explanans. This might be seen as some relevant piece of information, but the fact remains that a scientifically satisfactory explanation should mention much more then just the mere fact that \eqref{efe} hold --for example, if \eqref{efe} are cast in Hamiltonian form in a way that entails a $3+1$ decomposition of the dynamics, then some boundary conditions together with the specification of the initial conditions on a spacelike Cauchy surface would surely enter a satisfactory explanation of the dynamical development.

Hence, it is doubtful whether Livanios' relation of conceptual/nomic necessity can be enough to do any interesting explanatory job. If that is all that the anti-causal party can say about dependencies in general relativity, then they do not seem to fare any better than the friends of causation in the debate. 

At this point, if we want to make some progress, we need to pinpoint some minimal desiderata that a convincing non-causal characterization of the dependence relation underlying \eqref{efe} should fulfill, and then see if metaphysicians have something interesting to propose along these lines.

The first two desiderata should be basically a non-causal counterpart of the prominent features of causal dependence. Hence, the first requirement is that such a relation has to have explanatory power: for example, it should provide a conceptually robust backbone for explaining inertial motions via geometrical structure. The second desideratum is that it has to be a relation of determination in a strong ontological sense, that is, the existence of a condition determining the existence of another (e.g., the fact that freely-falling bodies move in such and such a way obtains in virtue of the fact that spacetime has such and such geometry). However, to avoid the problem that causal relations face in general relativity, we should further require this relation not to be inherently modal, that is, not to require modal notions in order to be defined (so supervenience would not do for our purposes). Finally, in order to avoid triviality, we should exclude that such a relation is \emph{just} conceptual necessity.

After a moment of reflection, we realize that there might be a good fit for this description. Indeed, the above desiderata are fulfilled by a dependence relation that metaphysicians call \emph{metaphysical grounding}. The notion of grounding is a lively debated topic among metaphysicians and, for obvious reasons, we cannot go through the debate here (see the essays in \citealp{559} to have an idea of the state-of-the-art on the subject). For this reason, we will pick a particular view on grounding, nicely exemplified by Paul Audi:
\begin{quote}
Grounding is the relation expressed by certain uses of the phrase ``in virtue of ,'' as in ``the act is wrong in virtue of its non-moral properties.'' [...] We should not use ``in virtue of'' where it might express a reflexive relation, such as identity. Since grounding is a relation of determination, and closely linked to the concept of explanation, it is irreflexive and asymmetric. [...] On my view, grounding is a singular relation between facts, understood as things having properties and standing in relations. Facts, on this conception, are not true propositions, but obtaining states of affairs.\\
(\citealp{560}, pp. 102-103)
\end{quote}

This brief characterization of grounding might make sense in general relativity: the Einstein field equations describe how a set of material (or spatiotemporal) facts obtain \emph{in virtue of} a set of spatiotemporal (or material) facts obtaining. The word ``obtaining'' here has to be understood as \emph{ontological} rather than simply \emph{conceptual} determination. In this vein, grounding can be thought of as a partial ordering with respect of fundamentality, while causality is a partial ordering with respect to time. This feature sounds good, because it means that grounding is better-suited than causation for ordering facts \emph{about} time, which are encoded in a $4$-dimensional geometry.

So, have we finally selected a good candidate for the role of dependence relation is general relativity? Of course not!

There are at least two problems with metaphysical grounding as the relation underpinning the dependence described by \eqref{efe}. The first is that, on this view, Misner, Thorne, and Wheeler's motto becomes at best an awkward metaphor. The problem is not stylistic, but substantial: nobody would contend that there is something straightforwardly \emph{physical} in the way spacetime and matter act on each other, and the notion of metaphysical grounding does not seem to capture such a physical character in any way. For sure, nobody would claim that the slingshot effect clearly exemplifies a case of metaphysical grounding without feeling a bit embarrassed. The second --and even worse-- problem is that, according to \eqref{efe}, material facts depend on spatiotemporal ones \emph{and viceversa}. Hence, the requirement of asymmetry for grounding is in stark tension with the (prima facie) mutual dependence encoded in the Einstein field equations. Otherwise said, the laws of general relativity do not favor any claim regarding geometrical facts being more fundamental than material ones or the other way round. One may try to resist this conclusion by pointing out that the stress-energy tensor in \eqref{efe} depends on the metric tensor, thus claiming that it is geometric facts that ground material facts in a clear formal sense, and not viceversa. However, that would be too hasty: while it is true that the stress-energy tensor depends on the metric tensor, we should not overlook the fact that it also depends on material fields. Hence, the most we can say is that the stress-energy tensor represents some sort of relational property shared by geometry and material fields, but this does not make any of the two more fundamental than the other (cf. \citealp{37} for a nice historical and philosophical discussion of this point). 

We have finally reached the core of the issue at stake. According to the positions reviewed in this section and the previous one, it is quite uncontroversial that the Einstein field equations encompass a dependence relation between material and spatiotemporal facts. However, nobody so far has been able to come up with a convincing characterization of the dependence involved. For sure, such a relation is explanatory in nature, it is not identity (unless \eqref{efe} are taken to \emph{define} a stress-energy tensor, which is controversial to say the least), and it is not inherently modal (although it might have modal consequences, depending on the particular view of modality adopted). However,  it is too ``physics-friendly'' to be just grounding yet not ``physically active'' enough to be just causation. 

My take on the problem is that general relativity is trying to teach us a philosophical lesson here, namely, that we are drawing too sharp a line between grounding and causation. In the next section I am going to make this intuition clearer by highlighting the well-known structural similarities between grounding and causation. In particular, I will draw on a recent proposal to analyze both grounding and causation via a generalized formal framework designed to ``spot'' dependence relations among facts. 

\section{A unified analysis of grounding and causation}\label{3}

As already said, the conceptual similarities between grounding and standard causation have not gone unnoticed in the philosophical literature. To begin with, they are both usually modeled as irreflexive, transitive and, hence, asymmetric binary relations i.e. strict partial orderings. Moreover, it can be easily argued that they both take facts as relata. In a nutshell, talk of causation between events can be reduced to talk of causation between facts about events happening  \citep[see][for an extensive articulation and defense of this view]{506}. If we accept this similarity between the two, then it is easy to extend this analogy to the distinction between partial/full ground on the one hand, and  contributory/sufficient cause on the other. But the structural similarities between grounding and causation run deeper than this. This fact becomes apparent if we analyze textbook cases of both of them through the structural equation modeling (SEM) framework. SEM is an umbrella term that groups together many mathematical and statistical tools used to investigate possible ordered structures underlying huge datasets (see \citealp{513}, for a historical survey of this rather heterogeneous set of methodologies). For our purposes, we are interested in the particular framework for testing causal links originally developed in \citet{501}. As it will become apparent in a moment, such a framework shares some characteristic traits with the ``generic'' interventionist approach advocated by Bartels and discussed in section \ref{2}. However, the SEM framework shows some new interesting features that makes it more suitable for being applied in the context of general relativity.

Without going too much into technical details, a structural equation model for causation consists in a set of variables that can be further grouped into \emph{endogenous} (i.e. those variables modeling the dependent conditions in the causal chain), and \emph{exogenous} (those variables that bring about the effects in the causal chain). Exogenous and endogenous variables are related by a set of equations that give a quantitative description of the dependence relations involved. These equations are basically functional relations, where the independent variables are exogenous and the dependent variables are endogenous. There is no a priori constraint on the type of functions involved, the only desiderata being that (i) the model has to adequately reproduce the correlation patterns observed in the original dataset, and (ii) the causal chain has to be consistent with the general principles used to justify the model. It is important to note that, since the framework is able to describe structured dependence chains, the exogenous/endogenous distinction is relative: given a certain link of the chain, all variables ``downstream'' will be endogenous and all those ``upstream'' will be exogenous. To fix the ideas and keep things simple, in the following we will mainly discuss toy cases of single-link chains, so that the exogenous/endogenous distinction becomes absolute. Once the way the framework works becomes clear, generalization to more complex chains is straightforward.

How can we say that a model adequately reproduces the correlations? Very simply, we just assign to the exogenous variables the actual values of the dataset and we see if the functions posited give back the actual values of the endogenous variables. Once this is achieved, we test the consistency of the causal chain by wiggling each of exogenous variables (i.e. assigning to them values different from the actual ones) and see if the functional dependence determines an appropriate wiggling of the endogenous ones. This procedure becomes clearer if we rephrase it in terms of manipulationist counterfactuals: ``if an intervention were to set an exogenous variable to such and such value, the endogenous variables would have taken on such and such values''.\footnote{Also in this case, in order to avoid confounding cases, it is required that an intervention on a variable disrupts all the links upstream of this variable.} If the counterfactual pattern generated in this way is consistent with the particular laws underlying the process under scrutiny, then the model can be considered as successfully catching a causal structure underpinning the correlations observed in the dataset.

The standard toy example that shows how the SEM framework ``spots'' causation involves the correlation between the shattering of a window on the throwing of a stone. In this case, if we take the laws of classical mechanics to be true, we would guess that the dependent part of the system is the window. We thus use an endogenous variable ``$S$'' to model the fact that the window shatters and we let this variable take on two values $\{0,1\}$ depending on whether this fact obtains or not. Likewise, we employ the exogenous variable $T\in\{0,1\}$ to model whether the fact that the stone is thrown obtains or not. The functional dependence in this case would be simply $S\overset{\leftarrow}{=}T$. Notice that the symbol ``$\overset{\leftarrow}{=}$'' highlights that the relation between the two variables is asymmetric.

Imagine that our dataset concerns one hundred repetitions of an experiment consisting in throwing a stone of fixed mass, shape, and initial velocity against a standardly crafted window, showing that the window shatters in $99\%$ of the cases. Then our assignment of the actual value of the exogenous variable would be $T=1$, which gives $S=1$. So the model adequately reproduces the correlations in the dataset.\footnote{The model can be easily expanded to account for more complex situations. For example, if there are two stone throwers, then we have to add another exogenous variable $T'\in\{0,1\}$ and another structural equation, i.e. $S\overset{\leftarrow}{=}T'$. This also shows how the framework deals with cases of causal overdetermination.} The next step would then be to wiggle $T$ and see if the ensuing counterfactual patterns are consistent with the laws of materials science. We get that ``were an intervention to prevent the stone from being thrown, the window would not have shattered'' is true under the model (setting $T=0$ leads to $S=0$), while the counterfactual ``were an intervention to prevent the stone from being thrown, the window would have shattered'' turns out to be false as required, together with the counterfactual ``were the stone to be thrown, the window would not have shattered''. Moreover, by construction, the model makes backtracking counterfactuals such as ``were a manipulation to shatter the window, the stone would have been thrown'' false, because an intervention on the window would sever the link that $S$ has with $T$ (which implies that any manipulation on $T$ does not affect $S$).

The result of our analysis is that our model works, so it can reliably be taken as representing a genuine causal dependence underlying the correlations in the starting dataset. Note how this treatment inherits all the advantages that a manipulationist account of causation has against the usual Lewisian analysis. First of all, the framework dispenses with the talk of possible worlds. Second, it avoids backtracking counterfactuals by construction. Third, it permits not only the analysis of token cases of causation (i.e. \emph{the} stone with such and such characteristics thrown at \emph{the} window crafted in such and such way), but also of type cases (the equation $S\overset{\leftarrow}{=}T$ is totally general and applies to all cases of a single stone being thrown at a window). Fourth, the analysis is quantitative in nature,\footnote{This means that we can create more complex models where variables do not have to be binary. For example, we can have a more accurate stone/window model with stone's momentum as the exogenous variable and glass' deformation as the endogenous one.} which permits a deeper understanding of the causal dynamics involved.

That being said, the SEM framework so sketched differs substantially from the manipulationist approach discussed in section \ref{2}. The most striking difference between the two can be tracked down --recalling Woodward's quotation on page \pageref{woo}-- to the theoretical basis for assessing ``claims about what will happen to $E$ under an intervention on $C$''. In the approach discussed in section \ref{2}, interventions were evaluated directly against the laws or principles of the particular theory (or set of theories) which constituted a viable theoretical basis for the context under consideration. In this way, such laws had a twofold role, namely, to provide a coherent notion of what a manipulation amounts to, and to constitute a formal machinery for evaluating the effects  of any given manipulation. To get a more vivid idea of how this works, let's return to the stone/window example. In this case, as already said, a viable theoretical framework to back up the interventionist analysis might be Newtonian mechanics and, in particular, the laws of inelastic scattering. Note how these laws give a clear physical description of the situation we are analyzing \emph{and} provide the formal machinery for the quantitative evaluation of concrete cases in which the stone is (or is not) thrown against the window.

In the SEM framework, on the other hand, such a formal machinery for quantitative evaluations is provided by the structural equations themselves. Now, while such equations surely have to be consistent with the laws and principles of the theoretical framework adopted to justify the model, they may or may not \emph{coincide} with the actual form of the laws or principles involved in justifying the adoption of the model, depending on the level of detail required by the analysis. Again, in the stone/window example, the equation $S\overset{\leftarrow}{=}T$ is not a physical description of inelastic scattering, but it models how some (physical) facts would bring about other facts were the laws of inelastic scattering to be true: If all we want to test is the systematic obtaining (or non-obtaining) of $S$ under the condition that $T$ obtains (or does not obtain), then the model is detailed enough. If, instead, we wanted a detailed analysis of, say, how far the glass debris fly in relation to the initial momentum of the stone, then we would use a much more complicated set of structural equations, which would coincide with the laws of Newtonian dynamics (or be a suitable approximation thereof).

This distinction between structural equations and underlying laws has a clear bearing on the constraints that an intervention has to obey. For, if the model coincided with the actual dynamical description of the modeled process, any intervention had to be directly encoded in the fundamental equations of the underlying theory. However, in the SEM framework there is no such restriction, the only constraints being that the model be logically consistent with the underlying laws or principles, and correctly reproduce the correlations in the dataset.

This fact makes it possible in many cases to come up with structural equations which feature variables whose range can include values associated with physically or even metaphysically impossible interventions. This liberal attitude towards the wiggling of the exogenous variables is not a problem, because interventions by themselves are to be seen as a conceptual stress test for the structure of the model: if such a structure reproduces the observed correlations and remains consistent with the underlying justificatory principles under any ``strain'', then the model is a solid one. If you want, you can picture a causal model as a system of connected springs inside the rigid box of the justificatory principles invoked: it provides extra conceptual degrees of freedom to an otherwise conceptually rigid apparatus. This peculiarity of the SEM analysis looks very encouraging in view of applying it to the general relativistic context. Indeed, the main problem with Bartels' manipulationist proposal was that evaluating counterfactuals directly against the laws of general relativity dramatically constrained the scope of manipulations and, consequently, the effectiveness of the dependence analysis. In the SEM framework, counterfactuals would be evaluated against a structural model, thus giving more room for analytical maneuver, while still remaining consistent with the laws of the theory. We will elaborate more extensively on this point in the next section.

At this point, however, the objection already presented in footnote \ref{foot} becomes all the more daunting. Let's repeat it here: the nature of a cause is deeply entangled with the nature of the associated manipulation; hence, physically or even metaphysically impossible interventions would make for conceptually obscure --if not ill-defined-- causal conditions. Quoting again Woodward:
\begin{quote}
[T]he claim that an asteroid impact caused the extinction of the dinosaurs can be understood within an interventionist framework as a claim about what would have happened to the dinosaurs if an intervention had occurred to prevent such an asteroid impact during the relevant time period. In this case we have both (i) a reasonably clear conception of what such an intervention would involve and (ii) principled ways of determining what would happen if such an intervention were to occur. By contrast, neither (i) nor (ii) hold if we are asked to consider hypothetical interventions that make it the case that $2+2\neq4$ or that the same object is at the same time both pure gold and pure aluminum or that transform human beings into houseflies. Causal claims that require for their explication claims about what would happen under such interventions [\dots] are thus unclear or at least have no
legitimate role in empirical inquiry.\\
(\citealp{563}, p. 224)
\end{quote}

A possible reply to this objection is that the characterization of causal conditions on the SEM framework shifts the accent from the nature of manipulations to the \emph{aptness} of the corresponding variables in modeling the particular situation analyzed. For example, we can minimally require that apt variables should take on values that represent distinct conditions, that distinct values of the same variable should represent mutually exclusive alternatives, and that variables represent enough conditions to capture the structure of the situation being analyzed (cf. \citealp[][section 1.3, and references therein]{562}, for a discussion of aptness constraints). The important point is that these aptness constraints make the nature of causal conditions and their change intelligible and meaningful to the particular context considered. For example, there is no particular problem in constructing a scientifically justified causal model that backs up a claim like ``If human beings suddenly turned into houseflies, urban environments would collapse'', even if no scientific theory would give you a basis for characterizing the change of a variable from ``human being'' to ``housefly''. Likewise, we can perfectly understand that a manipulation that sets the variable to, say, ``Vulcan'' instead of ``housefly'' would lead to opposite effects on urban environments (since we all agree that Vulcans are more rational than both houseflies and human beings). Of course, the skeptic might still point out that such an approach works only on a case-by-case basis, especially when we already have a fairly clear idea of where to look in order to spot a dependence. To this claim I will not reply here, being content to notice that the analysis of spacetime/matter-dependencies in general relativity is in fact one of these felicitous cases.

The liberal attitude towards interventions is a key assumption if we want to show that the SEM framework analyzes cases of grounding in the same way as it handles standard causation. Here I will mainly draw from \citet[][section 2]{471} and \citet[][section 5]{472}, in which such an extension of the framework to grounding is proposed and worked out. 

Consider now a textbook case of grounding, namely the existence of singleton Socrates being grounded in the existence of Socrates. Clearly, our intuition here is that the fact that ``singleton Socrates exists'' obtains is dependent on the fact that ``Socrates exists'' obtains. Let's call the endogenous variable modeling the former fact $S\in\{0,1\}$, and the exogenous variable modeling the latter fact $T\in\{0,1\}$. Not surprisingly, the structural equation modeling this dependence would be $S\overset{\leftarrow}{=}T$. The assignment of the actual value to $T$ would be $1$, which gives $S=1$: this models the truth of ``If Socrates exists, then singleton Socrates exists''. Now we wiggle $T$ to zero and, as a consequence, $S$ becomes zero. This symbolizes the truth of the counterfactual ``were an intervention to prevent Socrates from existing, singleton Socrates would not have existed'', which is exactly what we would have expected given our metaphysical treatment of grounding. 

Moreover, the model falsifies by construction the counterfactual ``were an intervention to prevent singleton Socrates from existing, Socrates would not have existed'' (see \citealp{503}, section 4, for a detailed analysis of how the SEM framework treats this counterfactual). Note that the antecedent of this counterfactual is metaphysically impossible. This is the case not because there are no metaphysically possible ways for singleton Socrates to fail to exist, but because, in this context, an external intervention that \emph{directly} sets $S$ to $0$ would sever the metaphysically necessary link between the ground $T$ and grounded $S$. Just to be clear, an intervention consisting in killing Socrates (or Socrates' mother) would not count as a direct (external) intervention on singleton Socrates, since it would in fact exploit the link from $T$ to $S$. If the reader is now having a very hard time thinking of a way to directly intervene on $S$ without passing through $T$, then she is starting to grasp what a metaphysically impossible intervention means in this context (cf. \citealp{472}, section 6, for a discussion of this type of impossible interventions). This, in a nutshell, is why this generalized SEM framework needs to allow for metaphysically impossible interventions.\footnote{To be fair, some authors (e.g. \citealp{564}) tend to deny that the obtaining of the grounding fact \emph{always} necessitates the obtaining of the grounded fact. However, in order to defend the need for metaphysically impossible interventions in the generalized SEM framework, we just need that necessitation between ground and grounded holds at least \emph{sometimes}.} 

If we agree on this analysis, then it is clear that the above statement depicts a \emph{counterpossible} situation on the considered model. Now, according to standard counterfactual semantics, counterpossible statements are trivially true, which is clearly unacceptable in this context. This is why this treatment demands a new, non-standard semantics to be built up (see, again, \citealp{503} for a preliminary discussion of this issue). Also in this case, the framework is able to account for both the token case involving Socrates, and the type case involving a singleton and its element.

Once the analogy between the stone/window and the element/singleton cases is acknowledged, it is easy to see how the framework can be pushed further to handle dependencies involving absences. For example, we can have $T\in\{0,1\}$ modeling the fact that I water the plant, and $S\in\{0,1\}$ modeling the fact that the plant dies. According to the laws that govern the physiology of plants, the structural equation in this case would be $S\overset{\leftarrow}{=}1-T$. Readers without a green thumb have certainly experienced the validity of this model. It is a simple exercise to show that this model works perfectly well also if we say that $T$ models the fact that unicorns exist and $S$ models the fact that the set of unicorns is empty. Of course, these analogies can be pushed even further, to embrace more complex models.

To sum up, if we generalize the SEM framework for causation to include a ``liberal'' view of manipulationist counterfactuals and a non-standard semantics for counterpossibles, then we obtain a powerful analytical tool to analyze not only instances of causation, but also instances of grounding. This generalized SEM framework is very peculiar in that, from a structural perspective, it is \emph{blind} to any distinction between grounding and causation. For example, it does not distinguish the stone/window case from the element/singleton case: from the framework's point of view they are both instances of a structural dependence of the form $X\overset{\leftarrow}{=}Y$. However, there is a clear sense in which, even in the generalized SEM framework, we can still distinguish grounding from causation. This distinction basically boils down to the ``mediating'' or ``formative'' principles used, that is, the set of principles invoked to justify the structural relations of a model. If these principles are just laws of nature (as in the stone/window case), then we are dealing with an instance of causation; if just metaphysical or mathematical principles are invoked (as in the element/singleton case), then that is a case of grounding (see \citealp{556} for a thorough defense of this law-based demarcation criterion against more usual criteria, such as that stating that causation is a diachronic relation, while grounding is not). Nevertheless, such a distinction is strictly speaking \emph{external} to the SEM framework, i.e., it is superimposed on the formalism without having any bearing on the results. From an internal point of view, what matters is just the functional dependencies established.

So, how far should we take the structural analogy between grounding and causation highlighted by the generalized SEM framework presented? The most conservative reaction is not to read too much into this. Even if grounding and causation might share some similar features, this is not enough to make the case for them to be related in a way that goes beyond some sort of vague conceptual resemblance (e.g. both being mentioned in explanation-related context), let alone considering them as metaphysically related (e.g. one being reducible to the other). However, nothing speaks against taking the grounding/causation analogy more seriously. In fact, there is a class of SEM models where, I believe, the idea of grounding and causation being somewhat metaphysically related gains traction. This class of models involves mixed chains of dependencies. 

To have a down-to-earth example, imagine a case in which a football player crosses the ball inside the penalty area and hits a defender's hand, thus determining a penalty kick. An appropriate functional model of this situation would be justified on the basis of both the laws of physics (the ball being kicked in such and such a way determines a trajectory that encounters the defender's hand at a certain point in space) and the rules of football (the handball determining a penalty kick). 

If we think that grounding and causation are substantially unrelated concepts, the SEM analysis of this situation looks a bit artificial in that there is a formal continuity in the treatment of the chain of dependencies that leads from crossing the ball into the penalty area to awarding a penalty kick yet a conceptual discontinuity involving some links of this chain. In other words, even if we depict the situation as a seamless chain of dependent facts --which, by the way, perfectly fits our intuitions--, from an ontological perspective, we nevertheless have two separate series of dependent facts, namely, (i) the crossing of the ball \emph{causing} the ball hitting the defender's hand, and (ii) the penalty kick being awarded \emph{in virtue of} the handball being sanctioned. 

If, instead, we are willing to accept that some sort of relation between causation and grounding really holds, then no such issue arises. The dependence chain regains its conceptual unity through the relatedness of the two concepts, and this also explains why the SEM analysis of mixed chains perfectly fits our everyday understanding of the situation as a seamless sequence of facts. This is indeed a key point to be highlighted since, as we are going to see in a moment, dependence chains in general relativity are in fact mixed.

\section{Dependence relations in general relativity: A possible answer}\label{4}
Let's step aside for a moment from the grounding/causation relatedness question, and let's just try to apply the generalized SEM analysis to the case of geodesic motions being dependent on geometrical facts in general relativity. As we are going to see, resorting to the SEM framework gives the issue a clear formal setting and, by doing so, brings to the fore the source of the confusion surrounding the debate reviewed in sections \ref{2} and \ref{2bis}.

Before doing this, however, let's see how the generalized SEM framework is able to address the challenges that standard analyses of dependencies face in the context of general relativity. The following discussion comes with an important caveat, namely, that the application of the SEM framework to concrete cases from general relativity is still a mostly unexplored line of research: to my knowledge, the only paper that follows this approach is \citet{557}, where the question regarding the causal nature of frame-dragging effects in general relativity is analyzed in terms of generalized structural models. As a consequence, what I am going to present here is the general conceptual strategy thanks to which the SEM framework deals with the issues related to dependence analysis. The concrete implementation of such a strategy would clearly depend on the particular case studied.

The first issue regards the way we can evaluate counterfactual changes. As we have seen, given a solution $\langle g_{\mu\nu},T_{\mu\nu}\rangle$ of \eqref{efe} representing the ``actual'' situation, there is in general no clear-cut criterion to single out the solution $\langle \tilde{g}_{\mu\nu},\widetilde{T}_{\mu\nu}\rangle$ that is ``most similar'' to $\langle g_{\mu\nu},T_{\mu\nu}\rangle$ and represents the implementation of the counterfactual change considered. To put it in the language of possible world semantics, there is in general no cogent way to single out the possible world nearest to the actual one because there is no spatiotemporal structure ``shared''  among possible worlds that constitutes a standard against which counterfactual variation can be evaluated. By adopting the SEM framework, we solve the issue because we do not evaluate the truth values of counterfactuals directly against the solutions of \eqref{efe}, but against a given structural model, i.e. by computing the values of the corresponding variables of the structural equations modeling a given situation. Using fancy words, we can say that it is the particular structural model chosen that determines the \emph{modal horizon} of a given situation, which means that counterfactual reasoning is structural model-dependent. For example, going back to the disappearing mass case of section \ref{2}, if the structural model we are working with is informative enough and rightly captures the dependencies in the actual world, we just have to set the material variables in the dependence chain to zero and check to see what geometric facts obtain as a consequence in order to see which vacuum solution of general relativity represents this new situation. If, on the other hand, by setting the material variables to zero we get a situation incompatible with the laws of general relativity (e.g. geometrical facts not encoded in any vacuum solution of \eqref{efe}), this means that the structural model does not pass the test, and so it is not accurate.

The problem of trans-world identification of spacetime points and regions is dealt with along the same lines. Under the SEM framework, any counterfactual change in a spacetime region $A$ in a solution of \eqref{efe} just translates into a change of value(s) of the variable(s) representing $A$ in the structural equations of an appropriate model. Again, if the structural model works well, this new situation will be depicted by another solution of \eqref{efe}, which will thus represent the counterfactual situation in which the very same $A$ were to be changed. Note how this way of defining identity by stipulation is akin to what physicists do in concrete cases. For example, when dealing with a perturbation of a background metric, they map the perturbed spacetime against the unperturbed background in a way that associates each point $P$ of the latter to a point $P'$ of the former (see, e.g., \citealp[][especially section 3]{568}). 

Likewise, for counterfactuals involving small local changes in a solution of \eqref{efe} that leave the rest of the universe untouched, all we have to do is to adopt an accurate, i.e. fine-grained enough, structural model that represents the actual dependence chain, ``zoom in'' the particular situation by looking at the variables that describe the local state of affairs, change their values as required, and then see whether and how this change globally affects the dependence chain. Clearly, we cannot perform such an operation directly on a solution of \eqref{efe}. Nonetheless, we can use the structural model to ``transition'' from the starting solution of the field equations (compatible with the starting values of the variables figuring in the structural equations) to a new one which encloses the changed state of affairs. The key requirement, once again, is that the structural model be fully compatible with the laws of general relativity, otherwise the changed state of affairs fixed by the new values of the structural variables might not be encoded in any solution of \eqref{efe}. Also in this case, the SEM framework offers a clear strategy to bypass the problem with model-wide interventions that Bartels' proposal had.

The reader unsympathetic towards this generalized SEM framework might point out that the general strategy sketched above does not eliminate the arbitrariness involved in the evaluation of counterfactuals, but just shifts it to a methodological or pragmatic perspective: if a structural model works, then we can use it for counterfactual reasoning without being bothered too much by metaphysical questions about modality. This is a fair remark which, however, does not challenge in any way the effectiveness of the SEM framework in treating counterfactual situations in general relativity. The general strategy I have just presented might be as conceptually nasty and inelegant as you want (depending on your philosophical tastes, of course), but still it provides a robust template for implementing structural models that are able to cope with the usual troubles that standard analyses of dependence face in general relativity.

Moving on, note how the SEM framework, in virtue of its loose and liberal interventionist attitude, does not require any notion of physical production in order to define the wiggling of the variables, thus dodging any potential issue related to the description of local physical processes in general relativity. If to this we add that the SEM framework deals with facts rather than physical events, we understand why it handles cases of causation by absence without being committed to the actual physical production of absences. Hence, if --for the sake of the argument-- we bite the bullet with Katzav's challenge and claim that spacetime just causes the absence of inexplicable changes in the trajectories of freely-falling bodies, we have no problem to coherently render such a claim via a structural model that resembles the plant watering case. Such a model would back up a perfectly fine explanation of why freely-falling bodies do not move erratically, which does not mention at any point that spacetime literally produces the absence of erratic motions.

Having argued that the SEM framework is a legitimate and promising approach to analyzing dependencies in general relativity, let us apply it to the case of geodesic motions. Our model aims to account for the geometry of spacetime $g_{\mu\nu}$ determining the set of geodesic motions $\left\{x^\mu_i(\lambda)\right\}_{i\in\mathbb{N}}$. In order to construct a general model of this dependence that is compatible with general relativistic dynamics, all we have to do is to follow the steps already sketched in section \ref{2} when discussing Bartels' proposal.

If we set the exogenous variables to model certain facts about the metric tensor itself and we link their variation to the endogenous variables encoding facts about the geodesic motions $x^\mu_i(\lambda)$ via a certain number of functional relations symbolized by $f_{i}$, in the end we get a set of structural equations that schematically look like $x^\mu_i(\lambda)\overset{\leftarrow}{=}f_i(g_{\mu\nu})$. To reiterate a point already made earlier, these equations need not be literally \eqref{efe}, \eqref{geo}, and \eqref{cris}. These latter equations would have to be used if we were interested in providing a detailed picture of the geodesic motions over a given spacetime, but here we want to construct a generic model that just captures the dependence of some facts on other facts in a theoretical context where the laws of general relativity hold. Hence, the role of \eqref{efe}, \eqref{geo}, and \eqref{cris} in this case is that of a conceptual constraint: for a structural model to rightfully capture the map of dependencies in a given situation, it has to be consistent with the laws of general relativity (hence, we cannot have situations where, say, by increasing spacetime curvature in a region $A$ the geodesic deviation decreases over $A$).

Given that we are interested in providing a general picture of how the analysis work, we are not interested in the particular form that these functional relations can take, the only constraint being that they capture, in the sense discussed above, the right counterfactual patterns encoded in general relativistic dynamics. So let's just assume that we get to a model with a chain of dependence of the form:
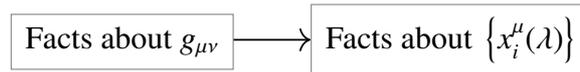
\begin{center}
\newcounter{dummy}
\refstepcounter{dummy}
\begin{tikzcd}[cells={nodes={draw=gray}}]\label{gga}
\text{Facts about } g_{\mu\nu} \arrow[r, black] & \text{Facts about } \left\{x^\mu_i(\lambda)\right\}
\end{tikzcd}
\end{center}
Is this dependence causal or metaphysical? This is a clear case in which standard grounding/causation demarcation criteria are ill-suited for the analysis. For example, the synchronicity/diachroneity criterion looks odd in a context where (i) there is no ``absolute'' temporal ordering against which we should evaluate the relationship between facts, and (ii) some facts are about time itself. The same limitations apply to the fundamentality/temporality criterion, especially given that the structural equations do not carry any useful information to evaluate the dependence based on this criterion. Fortunately, we can resort to the law-based demarcation criterion proposed at the end of section \ref{3}. According to this criterion, all we have to do is to look at the mediating principles that justify drawing the arrow in the above graph. By doing this, we immediately realize that the main justification comes from the Einstein field equations, from which the equations of motion \eqref{geo} follow. Given that the Einstein field equations are to be considered as laws of nature, it looks like we have reached a verdict: the above dependence is causal in nature. However, this conclusion would be too quick because we need another piece of information in order to justify such a dependence.

The problem is that the Einstein field equations do not single out a connection: many of them can go along with the same solution $\langle g_{\mu\nu}, T_{\mu\nu}\rangle$. This is why two constraints are placed ab initio on the connection, namely, (a) to be torsion-free, and (b) to be compatible with the metric tensor, i.e. to satisfy the relation $\boldsymbol{\nabla}\mathbf{g}=0$. It is important to stress the fact that these constraints are not placed to preserve the internal consistency of general relativity but, rather, they are theory-defining. In fact, relaxing or giving up on them would lead to entirely different but still consistent theories (see, e.g., \citealp{566}). Hence, (a) and (b) can be considered contingent law-like conditions rather than mathematical constraints that hold by metaphysical necessity.  These two constraints are enough to single out a unique connection because, according to the fundamental theorem of Riemannian geometry (whose proof can be found in any decent differential geometry textbook, such as \citealp{578}), there is just one torsion-free connection compatible with a given metric tensor, that is, the Levi-Civita connection (whose components are in fact given by \eqref{cris}). We thus have to sharpen our model in order to bring this point to the fore. To this extent, we add a further link in the chain:
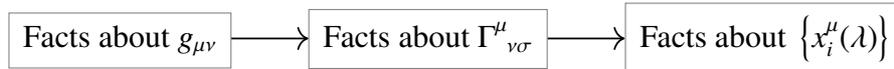
\begin{center}
\begin{tikzcd}[cells={nodes={draw=gray}}]
\text{Facts about } g_{\mu\nu} \arrow[r, black] & \text{Facts about } \Gamma^{\mu}_{\phantom{\mu}\nu\sigma} \arrow[r, black] & \text{Facts about } \left\{x^\mu_i(\lambda)\right\}
\end{tikzcd}
\end{center}
We immediately see that the law-like condition \eqref{geo} justifies the second link in the chain.\footnote{It is worth mentioning that the derivation of \eqref{geo} from \eqref{efe} through the condition that the covariant divergence of the stress-energy tensor vanishes works only if the Levi-Civita connection is used (thanks to an anonymous reviewer for suggesting this point to me).} Instead, the justification for the first link comes from constraints (a) and (b) --which can be considered part of the laws of general relativity-- \emph{and} the fundamental theorem of Riemannian geometry, which is a mathematical result. We hence have a case of mixed dependence. This conclusion nicely explains the roots of the confusion surrounding the debate reviewed in sections \ref{2} and \ref{2bis}: we simply cannot force a standard reading that is purely causal or ground-like on the dependence chain depicted above.

A possible objection to the analysis carried out so far is that it is totally misguided, since it presupposes a one-way determination relation from geometrical to material facts, where in fact the dependence relation that best depicts the Einstein field equations is \emph{mutual}: after all, this is exactly what the Misner, Thorne, and Wheeler motto is all about! To this I reply that, although it is true that there is no arrow on top of the equal sign in \eqref{efe}, still these equations do not depict a mutual dependence tout-court. All the Einstein field equations say is that there is no ``supremacy'' of material degrees of freedom over geometrical ones, and viceversa. Returning to the dependence chain depicted above, for sure we can render it more accurate by adding a further link upstream, thus having:  
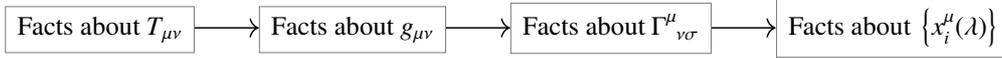
\begin{center}
\footnotesize
\begin{tikzcd}[cells={nodes={draw=gray}}]
\text{Facts about } T_{\mu\nu} \arrow[r, black] &\text{Facts about } g_{\mu\nu} \arrow[r, black] & \text{Facts about } \Gamma^{\mu}_{\phantom{\mu}\nu\sigma} \arrow[r, black]  & \text{Facts about } \left\{x^\mu_i(\lambda)\right\}
\end{tikzcd}
\end{center}
Now, the first link in this chain represents how material facts determine geometrical facts. However, there is no mutuality at all involved in this chain, since the material facts upstream are entirely different from the material facts downstream. For example, the first node may describe the overall material distribution of a binary system (e.g. on a Cauchy surface), while the last may describe the orbits that the system follows. 

The point becomes even stronger if we enlarge the upstream part of the chain, remembering the functional dependence of the stress-energy tensor:
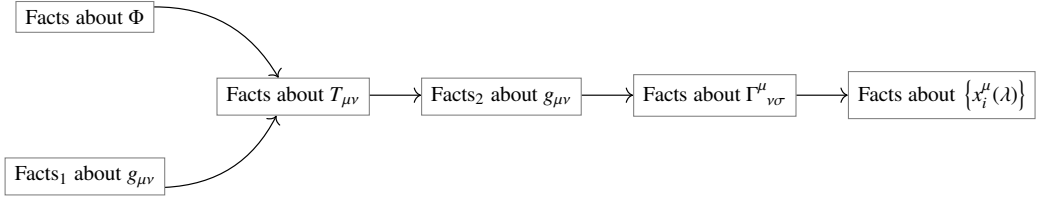
\begin{center}
\scriptsize
\begin{tikzcd}[cells={nodes={draw=gray}}]
 \text{Facts about }\Phi \arrow[dr, bend left, black] & & & &\\
 & \text{Facts about } T_{\mu\nu} \arrow[r, black] &\text{Facts$_2$ about } g_{\mu\nu} \arrow[r, black] & \text{Facts about } \Gamma^{\mu}_{\phantom{\mu}\nu\sigma} \arrow[r, black]  & \text{Facts about } \left\{x^\mu_i(\lambda)\right\}\\
\text{Facts$_1$ about } g_{\mu\nu} \arrow[ur, bend right, black] & & & &                      
\end{tikzcd}
\end{center}
Note that the lower first node does not coincide with the third and, likewise, the upper first node does not overlap with the second. Take, for example, the case of the Schwarzschild solution to \eqref{efe} in the interior of a star (see \citealp{100}, section 6.2, for a derivation of this solution). In this case ``Facts$_1$ about $g_{\mu\nu}$'' include that the geometry is static and spherically symmetric, while ``Facts$_2$ about $g_{\mu\nu}$'' involve a more detailed characterization of the geometry, including its dependence on the total mass of the body. Similarly, ``Facts about $\Phi$'' include that matter is a fluid with such and such density and pressure, while ``Facts about $T_{\mu\nu}$'' involve a detailed characterization of the stress-energy tensor, including the fact that the $4$-velocity of the fluid is such and such as a consequence of spacetime being static.

Coming back to a general discussion of the above chain, note that neither of the two first nodes is dependent on the other, which represents the fact that, according to general relativity, there is no material fact that is more fundamental than all geometric facts, and viceversa. In this sense, the analysis carried out in this paper is fully faithful to Misner, Thorne, and Wheeler motto (whose representation is in fact encoded in the above chain), and to the dependencies encompassed in \eqref{efe}. Moreover, note how, by expanding the dependence chain, its mixed character is reinforced. Indeed, now we have that the first (double) link is justified by (mathematical) functional dependence, while the second invokes the Einstein field equations plus boundary conditions.

In the end, what are the morals to be drawn from the above analysis? If we take seriously the generalized SEM framework, including the law-based demarcation criterion, then we have to conclude that the dependence relation underpinning general relativistic dynamics is definitely neither pure causation nor pure grounding.\footnote{Note that a similar conclusion would have been achieved even if we had adopted an alternative approach to the construction of general relativity, such as that of \citet{577}. According to this approach, the direction of the dependence depicted on page \pageref{gga} would have to be reversed. However, the analysis of the nature of the dependence involved would proceed similarly to the one presently carried out (the same mediating principles would have been invoked, only in a different order).} This, by itself, is not a huge achievement: already at the end of section \ref{2bis} we reached the conclusion that this relation is too ``physics-friendly'' to be just grounding yet not ``physically active'' enough to be just causation. However, the generalized SEM framework supplies us with a new perspective on the issue. Indeed, while before we assumed that grounding and causation were totally unrelated concepts, and this assumption made the conceptual tangle impossible to loosen, now the SEM framework --being a unified analytical framework for grounding and causation-- suggests to us that a way out of the impasse is in fact to accept that the two concepts are related in a substantial metaphysical sense, and general relativity is a theory that makes this relatedness manifest. This, of course, begs the question: in what way are grounding and causation related?

According to \citet{472}, they are just one and the same thing. Otherwise said, grounding is just a ``metaphysical'' way of causing. Hence, there is just a unique fundamental dependence relation --which is a conceptual primitive--, which may exhibit different ``flavors''. Wilson calls this fundamental relation ``causation'' and distinguishes the metaphysical flavor (which we would call ``grounding'') from the nomological one (which we would call ``standard causation''). However, this is just a matter of terminology. The gist of his thesis is that, where we previously saw two conceptually distinct dependence relations,  now there is in fact just \emph{one} dependence relation at work. The benefits of adopting this view are rather obvious: we get a parsimonious ideology and a unified explanatory framework with a relatively small effort. 

Under this view, there are no mixed chains of facts, since there is only one relation involved. Indeed, the distinction between laws of nature and metaphysical principles is useful only insofar as it explains why, up to now, metaphysicians had the \emph{impression} that grounding and causation were two conceptually distinct relations: simply, they have just focused on a distinction that cuts no metaphysical ice as far as the relation of determination is involved. Such a ``metaphysical illusion'' of there being two distinct dependence relations stems from the fact that metaphysicians usually focus on clear-cut, well-behaved everyday situations. When, however, fundamental physics comes into play, common intuitions are thrown out of the window, and the metaphysical unity of dependence relations becomes manifest, as the case of general relativity shows.

Wilson's thesis is rather strong and surely controversial (\citealp[see, e.g.,][section 4.5]{471}, for a critique of the grounding/causation unification thesis). However, we do not need to go that far to clarify the issue at stake. For example, we could maintain that, while grounding and causation are distinct concepts, still they are related in that they represent two determinates of a common determinable.\footnote{This idea is well-known in the literature. For example, \citet{565} argues that all the building relations are species of a unique genus relation.} Under this view, the law-based demarcation criterion determines a metaphysical scale of nomic dependence, causation being at one end (full nomic dependence) and grounding being at the other end (full non-nomic dependence). Therefore, the dependence relation underpinning \eqref{efe} falls somewhere in the middle of this scale, being some sort of metaphysical hybrid between causation and grounding. This explains why other analytical frameworks, which do not envisage this nomic dependence scale, are unfit to clarify the matter.

This is clearly not the place for adjudicating the dispute between dependence monists and pluralists. More modestly, the point that I want to drive home is that, by adopting a generalized SEM framework, it is possible (i) to elaborate a strategy to overcome the common conceptual troubles that standard analyses of dependence have in general relativity and (ii) to come up with some clear answers regarding the metaphysical status of the dependence relation encoded in the Einstein field equations. Of course, future work will have to show how the framework overcomes said conceptual troubles in concrete cases. Moreover, both monists and pluralists still have to deepen their own characterization of the discussed dependence relation --perhaps by looking at other fundamental physical theories. However, even at this earlier stage, I hope to have shown that the application of the generalized SEM framework to general relativity is a promising and exciting line of research, which intertwines physics with metaphysics.

\pdfbookmark[1]{Acknowledgements}{acknowledgements}
\begin{center}
\textbf{Acknowledgements}:
\end{center}
Many thanks to Carl Hoefer, Vassilis Livanios, Al Wilson, and two anonymous referees for their comments on earlier drafts of this paper. \c{C}a va sans dire, I am solely responsible for any remaining frown-inducing material. Also, I acknowledge financial support from the Spanish Ministry of Science, Innovation and Universities, fellowship IJCI-2015-23321.

\pdfbookmark[1]{References}{references}
\bibliography{biblio}

\begin{thebibliography}{}

\bibitem[\protect\citeauthoryear{Audi}{Audi}{2012}]{560}
Audi, P. (2012).
\newblock A clarification and defense of the notion of grounding.
\newblock In F.~Correia and B.~Schnieder (Eds.), {\em Metaphysical Grounding:
  {U}nderstanding the structure of reality}, Chapter~3, pp.\  101--121.
  Cambridge University Press.

\bibitem[\protect\citeauthoryear{Bartels}{Bartels}{2013}]{127}
Bartels, A. (2013).
\newblock Why metrical properties are not powers.
\newblock {\em Synthese\/}~{\em 190}, 2001--2013.

\bibitem[\protect\citeauthoryear{Bennett}{Bennett}{2011}]{565}
Bennett, K. (2011).
\newblock Construction area (no hard hat required).
\newblock {\em Philosophical Studies\/}~{\em 154}, 79--104.

\bibitem[\protect\citeauthoryear{Bird}{Bird}{2009}]{384}
Bird, A. (2009).
\newblock Structural properties revisited.
\newblock In T.~Handfield (Ed.), {\em Dispositions and causes}, Chapter~8, pp.\
   215--241. Oxford University Press.

\bibitem[\protect\citeauthoryear{Blanchard and Schaffer}{Blanchard and
  Schaffer}{2017}]{562}
Blanchard, T. and J.~Schaffer (2017).
\newblock Cause without default.
\newblock In H.~Beebee, C.~Hitchcock, and H.~Price (Eds.), {\em Making a
  difference}, Chapter~10, pp.\  175--214. Oxford University Press.

\bibitem[\protect\citeauthoryear{Brown and Lehmkuhl}{Brown and
  Lehmkuhl}{2016}]{438}
Brown, H. and D.~Lehmkuhl (2016).
\newblock Einstein, the reality of space, and the action-reaction principle.
\newblock In P.~Ghose (Ed.), {\em Einstein, Tagore and the Nature of Reality},
  Chapter~1, pp.\  9--36. Routledge.

\bibitem[\protect\citeauthoryear{Bruni, Matarrese, Mollerach, and Sonego}{Bruni
  et~al.}{1997}]{568}
Bruni, M., S.~Matarrese, S.~Mollerach, and S.~Sonego (1997).
\newblock Perturbations of spacetime: gauge transformations and gauge
  invariance at second order and beyond.
\newblock {\em Classical and Quantum Gravity\/}~{\em 14\/}(9), 2585--2606.

\bibitem[\protect\citeauthoryear{Correia and Schnieder}{Correia and
  Schnieder}{2012}]{559}
Correia, F. and B.~Schnieder (Eds.) (2012).
\newblock {\em Metaphysical Grounding: {U}nderstanding the structure of
  reality}.
\newblock Cambridge University Press.

\bibitem[\protect\citeauthoryear{Curiel}{Curiel}{2000}]{386}
Curiel, E. (2000).
\newblock The constraints general relativity places on physicalist accounts of
  causality.
\newblock {\em Theoria\/}~{\em 15\/}(1), 33--58.

\bibitem[\protect\citeauthoryear{Curiel}{Curiel}{2015}]{487}
Curiel, E. (2015).
\newblock If metrical structure were not dynamical, counterfactuals in general
  relativity would be easy.
\newblock \url{https://arxiv.org/abs/1509.03866}.

\bibitem[\protect\citeauthoryear{Dowe}{Dowe}{2000}]{475}
Dowe, P. (2000).
\newblock {\em Physical causation}.
\newblock Cambridge University Press.

\bibitem[\protect\citeauthoryear{Earman}{Earman}{1989}]{118}
Earman, J. (1989).
\newblock {\em World enough and space-time. Absolute versus relational theories
  of spacetime}.
\newblock The MIT Press.

\bibitem[\protect\citeauthoryear{Ehlers, Pirani, and Schild}{Ehlers
  et~al.}{1972}]{577}
Ehlers, J., F.~Pirani, and A.~Schild (1972).
\newblock The geometry of free fall and light propagation.
\newblock In L.~O'{R}eifeartaigh (Ed.), {\em General Relativity, papers in
  honour of {J. L. S}ynge}, pp.\  63--84. Clarendon Press.

\bibitem[\protect\citeauthoryear{Hehl, von~der Heyde, Kerlick, and Nester}{Hehl
  et~al.}{1976}]{566}
Hehl, F., P.~von~der Heyde, G.~Kerlick, and J.~Nester (1976).
\newblock General relativity with spin and torsion: {F}oundations and
  prospects.
\newblock {\em Reviews of modern physics\/}~{\em 48\/}(3), 393.

\bibitem[\protect\citeauthoryear{Hoefer}{Hoefer}{2014}]{504}
Hoefer, C. (2014).
\newblock Mach's principle as action-at-a-distance in {GR}: The causality
  question.
\newblock {\em Studies in History and Philosophy of Modern Physics\/}~{\em 48},
  128--136.

\bibitem[\protect\citeauthoryear{Jaramillo and Lam}{Jaramillo and
  Lam}{2018}]{574}
Jaramillo, J. and V.~Lam (2018).
\newblock Counterfactuals in the initial value formulation of general
  relativity.
\newblock {\em The {B}ritish Journal for the Philosophy of Science\/}~{\em
  axy066}.
\newblock {DOI} doi.10.1093/bjps/axy066.
  \url{http://philsci-archive.pitt.edu/15067/}.

\bibitem[\protect\citeauthoryear{Katzav}{Katzav}{2013}]{387}
Katzav, J. (2013).
\newblock Dispositions, causes, persistence as is, and general relativity.
\newblock {\em International studies in the philosophy of science\/}~{\em
  27\/}(1), 41--57.

\bibitem[\protect\citeauthoryear{Lam}{Lam}{2011}]{38}
Lam, V. (2011).
\newblock Gravitational and non-gravitational energy: the need for background
  structures.
\newblock {\em Philosophy of Science\/}~{\em 78}, 1012--1023.
\newblock \url{http://philsci-archive.pitt.edu/8372/}.

\bibitem[\protect\citeauthoryear{Lee}{Lee}{2009}]{578}
Lee, J. (2009).
\newblock {\em Manifolds and differential geometry}.
\newblock American Mathematical Society.

\bibitem[\protect\citeauthoryear{Lehmkuhl}{Lehmkuhl}{2008}]{50}
Lehmkuhl, D. (2008).
\newblock Is spacetime a gravitational field?
\newblock In D.~Dieks (Ed.), {\em The ontology of spacetime}, Volume~2 of {\em
  Philosophy and foundations of physics}, Chapter~5, pp.\  83--110. Elsevier
  B.V.

\bibitem[\protect\citeauthoryear{Lehmkuhl}{Lehmkuhl}{2011}]{37}
Lehmkuhl, D. (2011).
\newblock Mass-energy-momentum in general relativity. only there because of
  spacetime?
\newblock {\em British Journal for the Philosophy of Science\/}~{\em 62\/}(3),
  453--488.
\newblock \url{http://philsci-archive.pitt.edu/5137/}.

\bibitem[\protect\citeauthoryear{Leuenberger}{Leuenberger}{2014}]{564}
Leuenberger, S. (2014).
\newblock Grounding and necessity.
\newblock {\em Inquiry\/}~{\em 57\/}(2), 151--174.

\bibitem[\protect\citeauthoryear{Livanios}{Livanios}{2008}]{473}
Livanios, V. (2008).
\newblock Bird and the dispositional essentialist account of spatiotemporal
  relations.
\newblock {\em Journal for General Philosophy of Science\/}~{\em 39}, 383--394.

\bibitem[\protect\citeauthoryear{Livanios}{Livanios}{2017}]{512}
Livanios, V. (2017).
\newblock {\em Science in metaphysics}.
\newblock Palgrave Macmillan.

\bibitem[\protect\citeauthoryear{McKitrick}{McKitrick}{2005}]{505}
McKitrick, J. (2005).
\newblock Are dispositions causally relevant?
\newblock {\em Synthese\/}~{\em 144}, 357--371.

\bibitem[\protect\citeauthoryear{Mellor}{Mellor}{1995}]{506}
Mellor, H. (1995).
\newblock {\em The facts of causation}.
\newblock Routledge.

\bibitem[\protect\citeauthoryear{Misner, Thorne, and Wheeler}{Misner
  et~al.}{1973}]{27}
Misner, C., K.~Thorne, and J.~Wheeler (1973).
\newblock {\em Gravitation}.
\newblock W.H. Freeman and Company.

\bibitem[\protect\citeauthoryear{Pearl}{Pearl}{2000}]{501}
Pearl, J. (2000).
\newblock {\em Causality: Models, Reasoning, and Inference}.
\newblock Cambridge University Press.

\bibitem[\protect\citeauthoryear{Read}{Read}{2018}]{575}
Read, J. (2018).
\newblock Functional gravitational energy.
\newblock {\em The {B}ritish Journal for the Philosophy of Science\/}~{\em
  axx048}.
\newblock {DOI} 10.1093/bjps/axx048.

\bibitem[\protect\citeauthoryear{Schaffer}{Schaffer}{2016}]{471}
Schaffer, J. (2016).
\newblock Grounding in the image of causation.
\newblock {\em Philosophical Studies\/}~{\em 173}, 49--100.

\bibitem[\protect\citeauthoryear{Tamir}{Tamir}{2012}]{576}
Tamir, M. (2012).
\newblock Proving the principle: {T}aking geodesic dynamics too seriously in
  {E}instein's theory.
\newblock {\em Studies in History and Philosophy of Modern Physics\/}~{\em 43},
  137--154.

\bibitem[\protect\citeauthoryear{Vassallo}{Vassallo}{2016}]{467}
Vassallo, A. (2016).
\newblock A metaphysical reflection on the notion of background in modern
  spacetime physics.
\newblock In L.~Felline, A.~Ledda, F.~Paoli, and E.~Rossanese (Eds.), {\em New
  Directions in Logic and the Philosophy of Science}, pp.\  349--365. College
  Publications.
\newblock \url{http://arxiv.org/abs/1602.06254}.

\bibitem[\protect\citeauthoryear{Vassallo and Hoefer}{Vassallo and
  Hoefer}{2019}]{557}
Vassallo, A. and C.~Hoefer (2019).
\newblock The metaphysics of {M}achian frame-dragging.
\newblock In C.~Beisbart, T.~Sauer, and C.~W\"uthrich (Eds.), {\em Thinking
  about space and time}, Einstein Studies. Birkh\"auser.
\newblock \url{http://arxiv.org/abs/1901.10766}.

\bibitem[\protect\citeauthoryear{Wald}{Wald}{1984}]{100}
Wald, R. (1984).
\newblock {\em General Relativity}.
\newblock The University of Chicago press.

\bibitem[\protect\citeauthoryear{Westland}{Westland}{2015}]{513}
Westland, C. (2015).
\newblock {\em Structural Equation Models}.
\newblock Springer.

\bibitem[\protect\citeauthoryear{Wilson}{Wilson}{2017}]{472}
Wilson, A. (2017).
\newblock Metaphysical causation.
\newblock {\em No{\^u}s\/}.
\newblock {DOI} 10.1111/nous.12190.

\bibitem[\protect\citeauthoryear{Wilson}{Wilson}{2018}]{503}
Wilson, A. (2018).
\newblock Grounding entails counterpossible non-triviality.
\newblock {\em Philosophy and Phenomenological Research\/}~{\em 96\/}(3),
  716--728.

\bibitem[\protect\citeauthoryear{Wilson}{Wilson}{2019}]{556}
Wilson, A. (2019).
\newblock Classifying dependencies.
\newblock In D.~Glick, G.~Darby, and A.~Marmodoro (Eds.), {\em The Foundation
  of Reality: {F}undamentality, Space and Time}. Oxford University Press.

\bibitem[\protect\citeauthoryear{Woodward}{Woodward}{2003}]{561}
Woodward, J. (2003).
\newblock {\em Making things happen: {A} theory of causal explanation}.
\newblock Oxford University Press.

\bibitem[\protect\citeauthoryear{Woodward}{Woodward}{2008}]{563}
Woodward, J. (2008).
\newblock Mental causation and neural mechanisms.
\newblock In J.~Hohwy and J.~Kallestrup (Eds.), {\em Being reduced: {N}ew
  essays on reduction, explanation, and causation}, Chapter~12, pp.\  218--262.
  Oxford University Press.

\bibitem[\protect\citeauthoryear{Woodward}{Woodward}{2016}]{502}
Woodward, J. (2016).
\newblock Causation and manipulability.
\newblock {\em The Stanford Encyclopedia of Philosophy\/}.
\newblock
  \url{https://plato.stanford.edu/archives/win2016/entries/causation-mani/}.

\end{thebibliography}
\end{document}